\documentclass{article}

\usepackage{arxiv}
\pdfoutput=1

\usepackage[utf8]{inputenc} % allow utf-8 input
\usepackage[T1]{fontenc}    % use 8-bit T1 fonts
\usepackage{hyperref}       % hyperlinks
\usepackage{url}            % simple URL typesetting
\usepackage{booktabs}       % professional-quality tables
\usepackage{amsfonts}       % blackboard math symbols
\usepackage{nicefrac}       % compact symbols for 1/2, etc.
\usepackage{microtype}      % microtypography
\usepackage{lipsum}         % Can be removed after putting your text content
\usepackage{graphicx}
\usepackage[numbers]{natbib}
\usepackage{doi}
\usepackage{amsmath,amssymb}
\usepackage{cleveref}       % smart cross-referencing

\title{Hyperbolic phonon-plasmon polaritons in a hBN-graphene van der Waals structure}

\date{}

\newif\ifuniqueAffiliation
\ifuniqueAffiliation % Standard variant of author block
\author{ \href{https://orcid.org/0000-0000-0000-0000}{\includegraphics[scale=0.06]{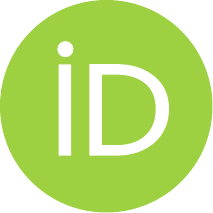}\hspace{1mm}Yu.~V.~Bludov}\thanks{Use footnote for providing further
		information about author (webpage, alternative
		address)---\emph{not} for acknowledging funding agencies.} \\
	Department of Physics, Center of Physics, and QuantaLab,\\
	University of Minho, Campus of Gualtar,\\
	4710-057, Braga, Portugal \\
	\texttt{hippo@cs.cranberry-lemon.edu} \\
	\And
	\href{https://orcid.org/0000-0000-0000-0000}{\includegraphics[scale=0.06]{orcid.pdf}\hspace{1mm}Elias D.~Striatum} \\
	Department of Electrical Engineering\\
	Mount-Sheikh University\\
	Santa Narimana, Levand \\
	\texttt{stariate@ee.mount-sheikh.edu} \\
}
\else
\usepackage{authblk}

\newbox{\orcid}\sbox{\orcid}{\includegraphics[scale=0.06]{orcid.pdf}} 
\author[1]{%
	\href{https://orcid.org/0000-0001-9648-1459}{\usebox{\orcid}\hspace{1mm}Yu.~V.~Bludov\thanks{\texttt{bludov@fisica.uminho.pt}}}%
}
\author[2,3]{%
	\href{https://orcid.org/0000-0003-3852-2085}{\usebox{\orcid}\hspace{1mm}D.~A.~Bahamon}%
}
\author[1,4,5]{%
	\href{https://orcid.org/0000-0002-7928-8005}{\usebox{\orcid}\hspace{1mm}N.~M.~R.~Peres}%
}
\author[2,3]{%
	\href{https://orcid.org/0000-0001-6165-3791}{\usebox{\orcid}\hspace{1mm}C.~J.~S.~de Matos}%
}
\affil[1]{Centro de Física (CF-UM-UP) e Departmento de Física, Universidade do Minho,
	PT-4710-057, Braga, Portugal}
\affil[2]{School of Engineering, Mackenzie Presbyterian University, São Paulo 01302-907, Brazil}
\affil[3]{MackGraphe-Graphene and Nanomaterials Research Institute, Mackenzie Presbyterian Institute, São Paulo 01302-907, Brazil}
\affil[4]{International Iberian Nanotechnology Laboratory
	(INL), Av. Mestre Jos\'{e} Veiga, 4715-330 Braga, Portugal}
\affil[5]{POLIMA---Center for Polariton-driven Light--Matter Interactions, University of Southern Denmark, Campusvej 55, DK-5230 Odense M, Denmark}
\fi

\renewcommand{\shorttitle}{Hyperbolic phonon-plasmon polaritons}

\hypersetup{
	pdftitle={Hyperbolic phonon-plasmon polaritons in a hBN-graphene van der Waals structure},
	pdfauthor={Yu.~V.~Bludov, D.~A.~Bahamon, N.~M.~R.~Peres, C.~J.~S.~de Matos},
}

\begin{document}
\maketitle

\begin{abstract}
In this paper a thorough theoretical study of a new class of collective excitations, dubbed hyperbolic surface phonon plasmon polaritons, is performed. This new type of light-matter excitations are shown to have unique properties that allows to explore them both as the basis of ultra-sensitive devices to the dielectric nature of its surroundings. The system is a van der Waals heterostructure -- a layered metamaterial, composed of different 2D materials in direct contact one with another, namely graphene ribbons and hexagonal boron nitride slabs of nanometric size. In the paper we discuss the spectrum of these new class of excitations, the associated electromagnetic fields, the sensitivity to the dielectric function of its surroundings, and the absorption spectrum. All this is accomplished using an analytical model that considerably diminishes the computational burden, as well as elucidates the underling physical mechanism of the excitations supported by the device.
\end{abstract}

\section{Introduction}

In solid state structures electromagnetic radiation can be effectively
coupled to the mechanical oscillations of charges, forming hybrid
modes, known as polaritons\cite{polaritons-review-Basov2020-nanoph}.
More specifically, electromagnetic radiation can be coupled to free charge
oscillations in metals (plasmons)\cite{plasmonics-review-Pitarke2007-RepProgPhys,plasmonics-review-Zhang2012-jpd,plasmonics-review-Anwar2018-dcn},
to phonons in dielectrics\cite{phonon-polariton-review-Caldwell2015-nanoph},
to excitons in semiconductor\cite{exciton-polariton-Kavokin2010-pssb},
or to the spin oscillations in magnetic materials (magnons)\cite{magnon-polaritons-Mills1974-rpp,magnon-polaritons-Camley1987-ssr}.
The resulting modes are known as plasmon-polariton, phonon-polariton,
exciton-polariton, and magnon-polariton, respectively. A general tendency
for the minituarization of electronic and photonic components attracted
researcher's interest to the investigation of the two-dimensional
solid-state structures, which are also able to sustain polaritons
with longer mean free path, if compared to the bulk structures. In
detail, surface plasmon-polaritons\cite{spp-gr-rev-Koppens2011-nl,spp-gr-rev-Xiao2016-fp,spp-gr-Nikitin2011-prb},
phonon-polaritons\cite{phonon-polariton-hBN-Dai2019-am} and exciton-polaritons\cite{exciton-polaritons-TMD-Hu2020-aom}
can exist in graphene, hexagonal Boron Nitride (hBN), and transition metal dichalcogenetes
(TMDs), correspondingly. At the same time, surface plasmon-polaritons
in graphene exist in the THz and far-infrared spectral range\cite{spp-gr-rev-Low2014-ACSNano},
i.e. in the domain of typical frequencies of optical phonons in dielectrics\cite{phonons-SiO2-Gunde2000-physb,phonons-SiO2-Zhang2012-prb,phonons-BN-Vuong2018-matmat,phonons-BN-Cai2007-ssc,phonons-BN-He2021-jmr,phonon-SiC-Koch2010-pss}.
The latter fact allows the interaction between plasmons, optical phonons
and electromagnetic radiation, which leads to the formation of hybrid
modes, known as surface phonon-plasmon polaritons\cite{spp-gr-phon-Luxmoore2014-acsphot,phonon-plasmon-polaritons-gr-Constant2016-natphys,spp-gr-SiC-Koch2010-prb}.
Experimentally these hybrid modes were observed in the graphene, deposited
on top of SiO$_2$\cite{spp-gr-phon-Fei2011-nl}, SiC\cite{spp-gr-phonon-sic-Liu2010-prb,spp-gr-SiC-Koch2010-prb}
or hBN\cite{spp-gr-phon-hBN-Brar2014-nl} substrates. When the thickness
of the hBN layer is finite, the spectrum of eigenmodes consists of infinite
number of waveguide modes\cite{phonon-polaritons-hBN-Low2017-natmat,phonon-polaritons-hBN-Dai2015-natcomm,phonon-polaritons-hBN-Caldwell2014-natcomm,phonon-polariton-hBN-Giles2018-matmat},
which exists in the upper and lower restrahlen bands (RBs), where,
respectively, longitudinal or transversal component of the dielectric
tensor is negative. 

\begin{figure}[t]
	\centering\includegraphics[width=8.5cm]{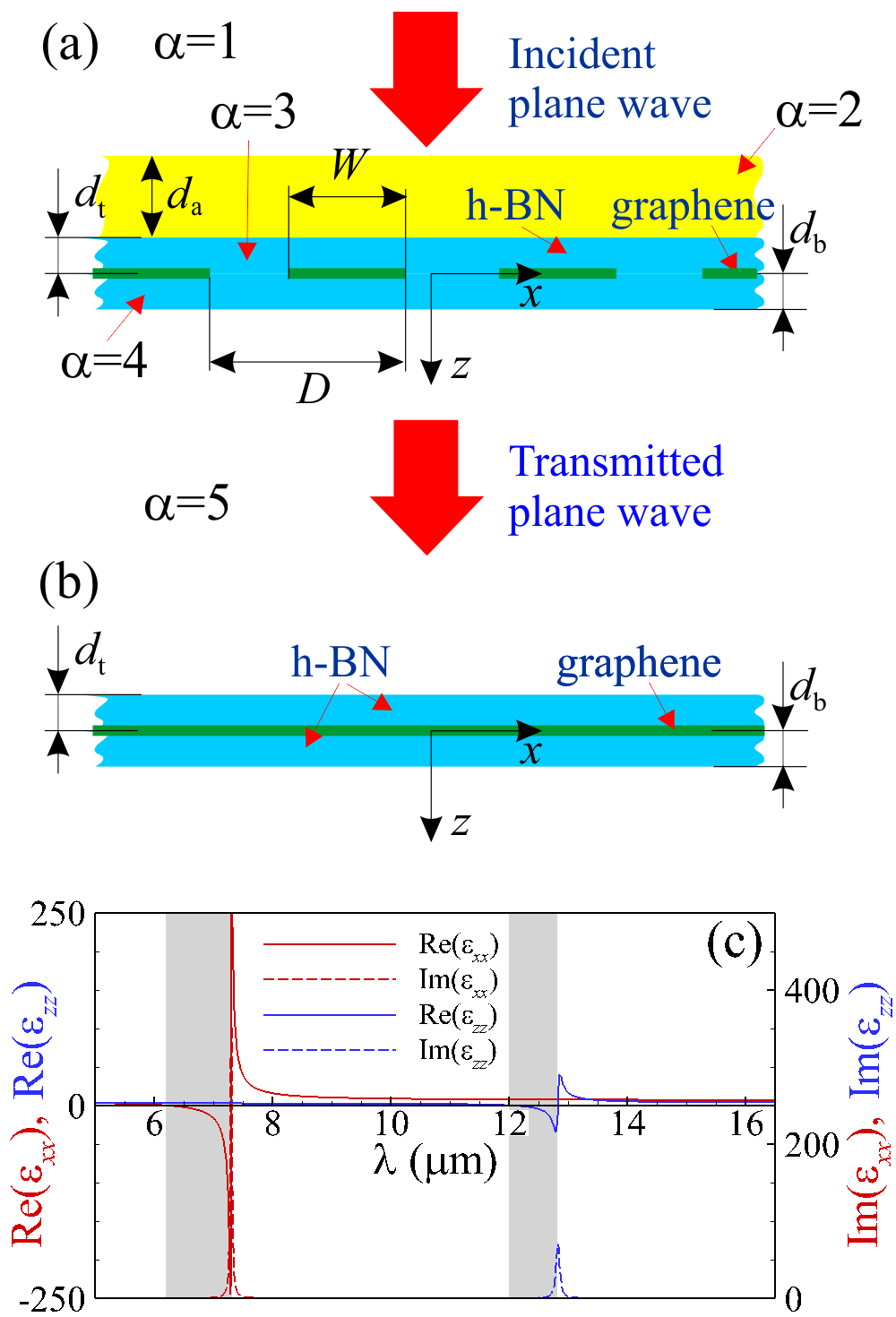}
	
	\caption{Layered structures, able to sustain surface phonon-plasmon polaritons:
		(a) periodic array of graphene nanostrips cladded by two layers
		of h-BN (with thicknesses of top and bottom layers $d_{t}$ and $d_{b}$,
		respectively), and covered with analyte layer of finite thickness;
		(b) Uniform graphene monolayer, cladded between two hBN layers; (c) Real (left axis) and imaginary (right axis) parts of the hBN dielectric tensor [see Eqs.\,(\ref{eq:eps_xx}) and (\ref{eq:eps_zz})]. Restrahlen bands are shadowed by grey color.}
	
	\label{fig:geometry}
\end{figure}

Since the modes mentioned above are surface modes (whose in-plane
wavevector is bigger than that of bulk electromagnetic wave in vacuum),
they can not be excited directly by an external wave, impinging on the
layered structure. One of the possibilities to overcome this wavevector
mismatch is to use the periodically patterned structure, where the excitation of hybrid mode takes place, when the wavevector of one of diffraction harmonics coincide with wavevector of eigenmode. Among other
possibilities, like nanoresonators\cite{gr-nanoresonators-Brar2013-nanolett,gr-split-rings-Sun2019-optcomm,gr-split-rings-Fan2018-acsphot,spp-gr-nanoresonators-nanoribbons-Rodrigo2016-prb,spp-gr-nanoresonators-Jang2014-prb},
disks\cite{gr-disks-exp-Yan2012-natnanotech,gr-disks-Fang2013-acsnano},
nanostructured substrate \cite{gr-subs-grating-Gao2013-nanolett,gr-subs-grating-Song2016-nrl,spp-gr-subs-grating-Song2016-acsphot,spp-gr-subs-grating-Qian2016-sensors}, or periodically patterned layer on top of graphene \cite{phonon-polaritons-MoO3-Yadav2021-oe,spp-gr-hBN-Hajian2019-jap},
a graphene monolayer patterned in form of nanoribbons\cite{gr-nanoribbons-Popov2010-prb,gr-nanoribbons-exp-Ju2011-nn,gr-nanoribbons-Hu2017-nanolett,gr-nanoribbons-Yan2013-natphot,gr-nanoribbons-photocurrent-Freitag2013-natcomm,gr-nanoribbons-Sorger2015-njp,gr-nanoribbons-Chu2013-apl}
attracts a special attention owing to both relatively easy sample
preparation and the possibility to theoretically model it, applying
simple analytical methods\cite{gr-nanoribbons-analytical-Goncalves2016-prb,gr-nanoribbons-analytical-Rahmanzadeh2018-josab,gr-nanoribbons-analytical-Zinenko2015-jopt}.
Another important application of an array of graphene nanoribbons is
their possibility to be used in sensing \cite{spp-gr-nanoribbons-sensor-Vasic2013-jap,spp-gr-nanoribbons-sensor-Hwang2021-scirep,spp-gr-nanoribbons-sensor-Bareza2023-ami,spp-gr-nanoribbons-sensor-Nong2020-achem,spp-gr-nanoribbons-sensor-Hu2019-natcomm,spp-gr-nanoribbons-sensor-Farmer2016-acsphot,spp-gr-nanoribbons-sensor-Li2014-nanolett,spp-gr-nanoribbons-sensor-Hu2018-advmat,spp-gr-nanoribbons-sensor-Rodrigo2015-science,spp-gr-nanoribbons-sensor-Wu2014-olt,spp-gr-nanoribbons-sensor-Hu2016-natcomm,spp-gr-nanoribbons-sensor-Souto2022-pssb}. Since in the sensing by means of graphene nanoribbons the analyte is arranged in the vicinity of graphene strips, where field of SPPs is maximal, phenomena in these structure have strong similarity with surface-enhanced infrared absorption (SEIRA), where weak light matter interaction is overcome by putting analyte into the hot spots of electromagnetic field\cite{sensor-seira-Miao2021-lsa,sensor-seira-Bareza2022-acsphot,sensor-seira-Yadav2023-ieeesens}.
In this paper a diffraction grating is realized by periodic patterning
the graphene. Thus, a graphene monolayer {[}located in $xy$-plane
at $z=0$, see Fig.\,\ref{fig:geometry}(a){]} is considered to be
patterned into an array of strips with period $D$. Each individual strip
is supposed to be infinite in the $y$-direction and of finite width $W$
in the $x$-direction {[}see Fig.\,\ref{fig:geometry}(a){]}, arranged
in the spatial interval $-W/2+lD\le x\le W/2+lD$, where $l$ is the
strip's number. The array of graphene nanostrips is cladded by two hBN layers of thicknesses $d_{b}$ and $d_{t}$ (for bottom
and top layers, respectively). Since this graphene-hBN hybrid structure
is supposed to be used in the plasmonic sensing application, the top
hBN layer is considered to be covered with an analyte, which is modelled
by the dielectric layer of thickness $d_{\mathrm{a}}$ and refractive
index $n_{\mathrm{a}}$, arranged in spatial domain $z_{2}=-d_{\mathrm{a}}-d_{t}\le z\le z_{3}=-d_{t}$.
The resulting layered structure is supposed to be truncated from both
sides by vacuum/air (for at $z>z_{5}=d_{b}$ and $z<z_{2}=-d_{\mathrm{a}}-d_{t}$).
We consider, that an external plane electromagnetic wave impinges on the
analyte from vacuum side.

\section{Diffraction of an electromagnetic wave by the periodic structure}

The fact that surface phonon-plasmon-polaritons are p-polarized waves
(where the direction of the magnetic field is perpendicular to the direction
of propagation), and the uniformity of the structures in Fig.\,\ref{fig:geometry}
along the $y$-direction, allows the description of their electromagnetic
properties by means of Maxwell's equations, which involves the $y$-component
of magnetic field, $\mathbf{H}=\left(0,H_{y},0\right)$, as well as
$x$- and $z$-components of the electric field, $\mathbf{E}=\left(E_{x},0,E_{z}\right)$.
Considering the electromagnetic field time-dependency as $\mathbf{E},\mathbf{H}\sim\exp{(-i\omega t)}$,
respective Maxwell's equations can be written in Gaussian units as
\begin{eqnarray}
\frac{\partial E_{x}^{(\alpha)}}{\partial z}-\frac{\partial E_{z}^{(\alpha)}}{\partial x}=\frac{i\omega}{c}H_{y}^{(\alpha)},\label{eq:max_Hy}\\
-\frac{\partial H_{y}^{(\alpha)}}{\partial z}=-\frac{i\omega}{c}\varepsilon_{xx}^{(\alpha)}E_{x}^{(\alpha)},\label{eq:max_Ex}\\
\frac{\partial H_{y}^{(\alpha)}}{\partial x}=-\frac{i\omega}{c}\varepsilon_{zz}^{(\alpha)}E_{z}^{(\alpha)}.\label{eq:max_Ez}
\end{eqnarray}
where $\omega$ is the wave cyclic frequency, $c$ is the velocity of
light in vacuum. Superscript $\alpha$ corresponds to the electromagnetic
fields in different dielectric layers, characterized by the dielectric
permittivity tensor, involving diagonal components only, i.e. $\varepsilon_{xx}^{(\alpha)}$
and $\varepsilon_{zz}^{(\alpha)}$.

Taking into account normal incidence of external electromagnetic wave, and, as a consequence, independence of electromagnetic field upon x-coordinate,
the solution of Maxwell's equations (\ref{eq:max_Hy})--(\ref{eq:max_Ez})
can be obtained in the form of a Fourier-Floquet series with respect
to cosine-like functions, which even symmetries coincide with that of incident wave. Sine-like harmonics can not be excited by normally incident wave owing to their odd symmetries. In the semi-infinite vacuum/air region $z<z_{2}=-d_{a}-d_{t}$
the solution of Maxwell's equations can be represented as a superposition
of forward-propagating (in $z$-direction) incident wave with amplitude
of magnetic field $H_{y}^{(i)}$, and infinite number of backward-propagating
spatial harmonics of reflected field with magnetic-field amplitudes $H_{y||m}^{(r)}$.
In the matrix form this type of solution can be represented as 
\begin{eqnarray}
\left(\begin{array}{c}
H_{y}^{(1)}(x,z)\\
E_{x}^{(1)}(x,z)
\end{array}\right)=\sum_{m=0}^{\infty}\cos\left[\frac{2\pi m}{D}x\right]\hat{F}_{m}\label{eq:HEv-vacuum}\\
\times\left(\begin{array}{c}
H_{y}^{(i)}\delta_{m,0}\exp\left[ip_{m}\left(z-z_{2}\right)\right]\\
H_{y||m}^{(r)}\exp\left[-ip_{m}\left(z-z_{2}\right)\right]
\end{array}\right),\nonumber 
\end{eqnarray}
where 
\begin{eqnarray*}
	\hat{F}_{m}=\left(\begin{array}{cc}
		1 & 1\\
		\frac{cp_{m}}{\omega} & -\frac{cp_{m}}{\omega}
	\end{array}\right)
\end{eqnarray*}
is the field matrix, 
\begin{eqnarray*}
	p_{m}=\sqrt{\left(\frac{\omega}{c}\right)^{2}-\left(\frac{2\pi m}{D}\right)^{2}}
\end{eqnarray*}
is the out-of-plane wavevector component of the $m$th harmonics in vacuum
(the respective in-plane component is equal to $2\pi m/D$), and $\delta_{m,m^\prime}$ is Kronecker delta. For
the other semi-infinite vacuum/air region, $z>z_{5}=d_{b}$, the solution
of Maxwell's equation includes forward-propagating harmonics of transmitted wave
only, i.e. 
\begin{eqnarray}
\left(\begin{array}{c}
H_{y}^{(5)}(x,z)\\
E_{x}^{(5)}(x,z)
\end{array}\right)=\sum_{m=0}^{\infty}\cos\left[\frac{2\pi m}{D}x\right]\hat{F}_{m}\label{eq:HEs-substrate}\\
\times\left(\begin{array}{c}
H_{y||m}^{(t)}\exp\left[ip_{m}\left(z-z_{5}\right)\right]\\
0
\end{array}\right).\nonumber 
\end{eqnarray}
Here zero in the second line means absence of the backward-propagating
harmonics, mentioned above. 

In the semi-infinite regions, occupied by air/vacuum, solutions
of Maxwell's equations [see Eqs.\,\eqref{eq:HEv-vacuum} and \eqref{eq:HEs-substrate}]
are parametrized by the amplitudes of the magnetic field of forward- and backward-propagating
harmonics. Inside the layers of finite thickness the solution of Maxwell's
equations will be parametrized in different manner -- by the amplitudes
of the electric- and magnetic fields' tangential components, $e_{x||m}^{(\alpha)}$
and $h_{y||m}^{(\alpha)}$, at one of the layer boundaries. The spatial
dependence of the electromagnetic field inside the dielectric slab is
determined by the transfer-matrix
\begin{eqnarray}
\hat{Q}_{m}^{(\alpha)}\left(z\right)=\left(\begin{array}{cc}
\cos\left[q_{m}^{(\alpha)}z\right] & \frac{i\omega\varepsilon_{xx}^{(\alpha)}\left(\omega\right)}{cq_{m}^{(\alpha)}}\sin\left[q_{m}^{(\alpha)}z\right]\\
\frac{icp_{m}^{(\alpha)}}{\omega\varepsilon_{xx}^{(\alpha)}\left(\omega\right)}\sin\left[q_{m}^{(\alpha)}z\right] & \cos\left[q_{m}^{(\alpha)}z\right]
\end{array}\right),\label{eq:Qm}
\end{eqnarray}
where in the case of anisotropic medium the out-of-plane component
of wavevector is represented as $q_{m}^{(\alpha)}=\sqrt{\left(\frac{\omega}{c}\right)^{2}\varepsilon_{xx}^{(\alpha)}\left(\omega\right)-\left(\frac{2\pi m}{D}\right)^{2}\varepsilon_{xx}^{(\alpha)}\left(\omega\right)/\varepsilon_{zz}^{(\alpha)}\left(\omega\right)}$.
Inside the top ($\alpha=3$) and bottom ($\alpha=4$) hBN layers,
the spatial dependence of total electromagnetic field have the form
\begin{eqnarray}
\left(\begin{array}{c}
H_{y}^{(3)}(x,z)\\
E_{x}^{(3)}(x,z)
\end{array}\right)=\sum_{m=0}^{\infty}\cos\left[\frac{2\pi m}{D}x\right]\nonumber \\
\times\hat{Q}_{m}^{(3)}\left(z+d_{t}\right)\left(\begin{array}{c}
h_{y||m}^{(3)}\left(z_{3}\right)\\
e_{x||m}^{(3)}\left(z_{3}\right)
\end{array}\right),\label{eq:HE-top-hBN}\\
\left(\begin{array}{c}
H_{y}^{(4)}(x,z)\\
E_{x}^{(4)}(x,z)
\end{array}\right)=\sum_{m=0}^{\infty}\cos\left[\frac{2\pi m}{D}x\right]\nonumber \\
\times\hat{Q}_{m}^{(4)}\left(z-d_{b}\right)\left(\begin{array}{c}
h_{y||m}^{(4)}\left(z_{5}\right)\\
e_{x||m}^{(4)}\left(z_{5}\right)
\end{array}\right).\label{eq:HE-bottom-hBN}
\end{eqnarray}
Here the components of hBN dielectric permeability tensor are\cite{spp-hbn-Kumar2015-nl}
\begin{eqnarray}
\varepsilon_{xx}^{(3)}(\omega)=\varepsilon_{xx}^{(4)}(\omega)=\varepsilon_{xx}\left(\infty\right)\nonumber\\
\times\left(1+\frac{\left(\omega_{xx}^{(LO)}\right)^{2}-\left(\omega_{xx}^{(TO)}\right)^{2}}{\left(\omega_{xx}^{(TO)}\right)^{2}-\omega^{2}-i\omega\Gamma_{xx}^{(TO)}}\right),\label{eq:eps_xx}\\
\varepsilon_{zz}^{(3)}(\omega)=\varepsilon_{zz}^{(4)}(\omega)=\varepsilon_{zz}\left(\infty\right)\nonumber\\
\times\left(1+\frac{\left(\omega_{zz}^{(LO)}\right){}^{2}-\left(\omega_{zz}^{(TO)}\right){}^{2}}{\left(\omega_{zz}^{(TO)}\right){}^{2}-\omega^{2}-i\omega\Gamma_{zz}^{(TO)}}\right),\label{eq:eps_zz}
\end{eqnarray}
with parameters $\omega_{xx}^{(TO)}=1370\thinspace\mathrm{cm^{-1}}$,
$\omega_{xx}^{(LO)}=1610\thinspace\mathrm{cm^{-1}}$, $\Gamma_{xx}^{(TO)}=5\thinspace\mathrm{cm^{-1}}$,
$\omega_{zz}^{(TO)}=780\thinspace\mathrm{cm^{-1}}$, $\omega_{zz}^{(LO)}=830\thinspace\mathrm{cm^{-1}}$,
$\Gamma_{zz}^{(TO)}=4\thinspace\mathrm{cm^{-1}}$, $\varepsilon_{xx}\left(\infty\right)=4.87$,
$\varepsilon_{zz}\left(\infty\right)=2.95$. The respective dependence of hBN dielectric permeability tensor components upon vacuum wavelength $\lambda$ are depicted in Fig.\,\ref{fig:geometry}(c). The analyte medium $\left(\alpha=2\right)$
is assumed to be isotropic throughout this paper, thus its dielecric
tensor components in Eq.\,(\ref{eq:Qm}) are assumed to be $\varepsilon_{xx}^{(2)}\left(\omega\right)=\varepsilon_{zz}^{(2)}\left(\omega\right)=n_{\mathrm{a}}^{2}$.
The respective electromagnetic fields inside the spatial domain, occupied
by the analyte layer, are
\begin{eqnarray}
\left(\begin{array}{c}
H_{y}^{(2)}(x,z)\\
E_{x}^{(2)}(x,z)
\end{array}\right)=\sum_{m=0}^{\infty}\cos\left[\frac{2\pi m}{D}x\right]\nonumber \\
\times\hat{Q}_{m}^{(2)}\left(z-z_{2}\right)\left(\begin{array}{c}
h_{y||m}^{(2)}\left(z_{2}\right)\\
e_{x||m}^{(2)}\left(z_{2}\right)
\end{array}\right).\label{eq:HE-analyte}
\end{eqnarray}

Since the boundary conditions at interfaces $z=z_{2},z_{3},z_{5}$
can be expressed as the continuity of the tangnetial components of
the electromagnetic fields, i.e. $\left(H_{y}^{(\alpha-1)}(x,z_{\alpha}),\quad E_{x}^{(\alpha-1)}(x,z_{\alpha})\right)^{T}=\left(H_{y}^{(\alpha)}(x,z_{\alpha}),\quad E_{x}^{(\alpha)}(x,z_{\alpha})\right)^{T}$,
the electromagnetic fields at both sides of graphene layer (arranged
at $z=z_{4}=0$) can be obtained by the multiplication of field- and
transfer matrices. Being expressed through amplitudes of the magnetic
field in the semi-infinite media, the respective fields take the form 
\begin{eqnarray}
\left(\begin{array}{c}
H_{y}^{(3)}(x,0)\\
E_{x}^{(3)}(x,0)
\end{array}\right)=\sum_{m=0}^{\infty}\cos\left[\frac{2\pi m}{D}x\right]\hat{F}_{m}^{(-,tot)}\label{eq:HE3-tot}\\
\times\left(\begin{array}{c}
H_{y}^{(i)}\delta_{m,0}\\
H_{y||m}^{(r)}
\end{array}\right),\nonumber \\
\left(\begin{array}{c}
H_{y}^{(4)}(x,0)\\
E_{x}^{(4)}(x,0)
\end{array}\right)=\sum_{m=0}^{\infty}\cos\left[\frac{2\pi m}{D}x\right]\hat{F}_{m}^{(+,tot)}\label{eq:HE4-tot}\\
\times\left(\begin{array}{c}
H_{y||m}^{(t)}\\
0
\end{array}\right),\nonumber 
\end{eqnarray}
where $\hat{F}_{m}^{(-,tot)}=\hat{Q}_{m}^{(3)}\left(d_{t}\right)\hat{Q}_{m}^{(2)}\left(d_{\mathrm{a}}\right)\hat{F}_{m}$,
$\hat{F}_{m}^{(+,tot)}=\hat{Q}_{m}^{(4)}\left(-d_{b}\right)\hat{F}_{m}$
are total field matrices above and below the graphene layer. The boundary
condition across the graphene has a different form: the electric field's
tangential component is continuous across the graphene, $E_{x}^{(4)}(x,0)=E_{x}^{(3)}(x,0)$,
while magnetic field is discontinuous owing to the presence of currents
in graphene, 
\begin{eqnarray}
H_{y}^{(4)}(x,0)-H_{y}^{(3)}(x,0)=-\left(4\pi/c\right)j_{x}\left(x\right).\label{eq:bc-H}
\end{eqnarray}
In the case of nanostrips array the current in graphene can be approximated
by square-root-like function \cite{gr-nanoribbons-analytical-Goncalves2016-prb}
\begin{eqnarray}
j_{x}\left(x\right)=\sum_{l=-\infty}^{\infty}j_{x}\left(0\right)\Theta\left(\frac{W}{2}-\left|x-lD\right|\right)\label{eq:current1}\\
\times\sqrt{1-\frac{4}{W^{2}}\left(x-lD\right)^{2}},\nonumber 
\end{eqnarray}
where $\Theta\left(x\right)$ is the Heaviside step-like function.
In this expression we applied Bloch theorem $j_{x}\left(lD\right)=j_{x}\left(0\right)$
and supposed, that at the edges of each individual strip the current
is zero, $j_{x}\left(\pm W/2+lD\right)\equiv0$, and there are no
currents between strips. Substututing magnetic fields from Eqs.\,\,(\ref{eq:HE3-tot}),
(\ref{eq:HE4-tot}), as well as current (\ref{eq:current1}) into
boundary condition (\ref{eq:bc-H}), we have
\begin{eqnarray}
\left(\hat{F}_{m}^{(+,tot)}\right)_{11}H_{y||m}^{(t)}-\left(\hat{F}_{m}^{(-,tot)}\right)_{11}H_{y}^{(i)}\delta_{m,0}\label{eq:bc-H-res}\\
-\left(\hat{F}_{m}^{(-,tot)}\right)_{12}H_{y||m}^{(r)}=-\frac{4\pi}{c}j_{x||m},\nonumber 
\end{eqnarray}
where 
\begin{eqnarray}
j_{x||m}=\frac{2\left(2-\delta_{m,0}\right)}{D}\intop_{0}^{W/2}j_{x}\left(x\right)\cos\left(\frac{2\pi m}{D}x\right)dx\label{eq:jxm}\\
=j_{x}\left(0\right)\frac{2-\delta_{m,0}}{2m}J_{1}\left(\frac{m\pi W}{D}\right)\nonumber 
\end{eqnarray}
is the Fourier component of current density, and $J_{1}\left(\cdot\right)$
is the Bessel function. In the similar manner, using boundary conditions
for electric field's tangential component, we obtain
\begin{eqnarray}
\left(\hat{F}_{m}^{(+,tot)}\right)_{21}H_{y||m}^{(t)}-\left(\hat{F}_{m}^{(-,tot)}\right)_{21}H_{y}^{(i)}\delta_{m,0}\label{eq:bc-E-res}\\
-\left(\hat{F}_{m}^{(-,tot)}\right)_{22}H_{y||m}^{(r)}=0.\nonumber 
\end{eqnarray}
From Eqs.\,(\ref{eq:bc-H-res}) and (\ref{eq:bc-E-res}) it is possible
to express amplitudes for the transmitted and reflected harmonics as
\begin{eqnarray}
H_{y||m}^{(r)}=-\frac{f_{0}}{g_{0}\left(\hat{F}_{0}^{(-,tot)}\right)_{22}\left(\hat{F}_{0}^{(+,tot)}\right)_{21}}H_{y}^{(i)}\delta_{m,0}\label{eq:Hyr-res}\\
-\frac{1}{g_{m}\left(\hat{F}_{m}^{(-,tot)}\right)_{22}}\frac{4\pi}{c}j_{x||m},\nonumber \\
H_{y||m}^{(t)}=-\frac{2}{g_{0}\left(\hat{F}_{0}^{(-,tot)}\right)_{22}\left(\hat{F}_{0}^{(+,tot)}\right)_{21}}H_{y}^{(i)}\delta_{m,0}\label{eq:Hyt-res}\\
-\frac{1}{g_{m}\left(\hat{F}_{m}^{(+,tot)}\right)_{21}}\frac{4\pi}{c}j_{x||m},\nonumber 
\end{eqnarray}
where $f_{0}=\left(\hat{F}_{0}^{(+,tot)}\right)_{11}\left(\hat{F}_{0}^{(-,tot)}\right)_{21}-\left(\hat{F}_{0}^{(-,tot)}\right)_{11}\left(\hat{F}_{0}^{(+,tot)}\right)_{21}$
and $g_{m}=\left(\hat{F}_{m}^{(+,tot)}\right)_{11}/\left(\hat{F}_{m}^{(+,tot)}\right)_{21}-\left(\hat{F}_{m}^{(-,tot)}\right)_{12}/\left(\hat{F}_{m}^{(-,tot)}\right)_{22}$.
On the other hand, current density in graphene strips can be expressed
via the Ohm law as
\begin{eqnarray*}
	j_{x}\left(x\right)=\sigma_{g}\left(\omega\right)\sum_{l=-\infty}^{\infty}\Theta\left(\frac{W}{2}-\left|x-lD\right|\right)E_{x}^{(4)}(x,0),
\end{eqnarray*}
where the graphene's conductivity $\sigma_{g}\left(\omega\right)$
is supposed to be of Drude type,
\begin{eqnarray*}
	\sigma_{g}\left(\omega\right)=\frac{e^{2}}{\hbar^{2}\pi}\frac{E_{F}}{\gamma-i\omega}.
\end{eqnarray*}
Here $E_{F}$ is the Fermi energy, $\gamma$ is the inverse relaxation
rate, $e$ is the electron charge, and $\hbar$ is the Planck constant.
In this case the average value of current on the period can be represented
in the form 
\begin{eqnarray*}
	j_{x||0}=\frac{2}{D}\intop_{0}^{D/2}j_{x}\left(x\right)dx\\
	=\sigma_{g}\left(\omega\right)\sum_{m=0}^{\infty}\frac{\sin\left(\frac{m\pi W}{D}\right)}{m\pi}\left(\hat{F}_{m}^{(+,tot)}\right)_{21}H_{y||m}^{(t)}.
\end{eqnarray*}
Combining this expression with the explicit form of $j_{x||m}$ {[}see
Eq.\,(\ref{eq:jxm}){]} and $H_{y||m}^{(t)}$ {[}see Eq.\,(\ref{eq:Hyt-res}){]},
it is possible to obtain an expression for current density at the
center of the strip,
\begin{eqnarray}
j_{x}\left(0\right)=-\sigma_{g}\left(\omega\right)\frac{8}{g_{0}\pi}\frac{H_{y}^{(i)}}{\left(\hat{F}_{0}^{(-,tot)}\right)_{22}}\left\{ 1+\frac{8}{c}\sigma_{g}\left(\omega\right)\right.\nonumber \\
\left.\times\sum_{m=0}^{\infty}\frac{D\sin\left(\frac{m\pi W}{D}\right)}{Wm\pi}\frac{1}{g_{m}}\frac{2-\delta_{m,0}}{m}J_{1}\left(\frac{m\pi W}{D}\right)\right\} ^{-1}.\label{eq:jx-at0}
\end{eqnarray}
This equation, jointly with Eqs.\,(\ref{eq:jxm}), (\ref{eq:Hyr-res}),
and (\ref{eq:Hyt-res}) allows calculation of the amplitudes of spatial
harmonics of the transmitted and reflected wave $H_{y||m}^{(t)}$,
$H_{y||m}^{(r)}$. For the limiting case of continuous graphene, $W=D$,
absence of analyte, and absence of both hBN layers {[}when total field
matrices are $\hat{F}_{m}^{(\pm,tot)}=\hat{F}_{m}${]} the current
density (\ref{eq:jx-at0}) will be represented in the simple form
\begin{eqnarray}
j_{x}\left(0\right)=\frac{\sigma_{g}\left(\omega\right)\frac{4}{\pi}H_{y}^{(i)}}{1+\frac{2\pi}{c}\sigma_{g}\left(\omega\right)}.\label{eq:jx-at0-1}
\end{eqnarray}
Respectively, the amplitudes of the reflected and transmitted waves
(zeroth harmonics) are
\begin{eqnarray*}
	H_{y||0}^{(r)}=\frac{\frac{2\pi}{c}\sigma_{g}\left(\omega\right)H_{y}^{(i)}}{1+\frac{2\pi}{c}\sigma_{g}\left(\omega\right)},\\
	H_{y||0}^{(t)}=\frac{H_{y}^{(i)}}{1+\frac{2\pi}{c}\sigma_{g}\left(\omega\right)},
\end{eqnarray*}
which coincide with the previous results for the case of continuous
graphene\cite{rt-graph-Bludov2013-Jopt}. 

Further in the paper the electromagnetic properties of the structure
will be characterized by the reflectance, transmittance, and absorbance,
which are related to the energy flux, whose absolute value and direction
are described by the Poynting vector $\mathbf{S}=\left(c/8\pi\right)\mathrm{Re}\left[\mathbf{E}\times\overline{\mathbf{H}}\right]$,
where overbar stands for complex conjugation, and $\times$ stands
for the vector product. Since incident wave propagates along $z$-axis, integral
reflection (transmission) coefficient is proportional to $z$-component
of the reflected (transmitted) wave's Poynting vector, averaged on
half-period, 
\begin{eqnarray}
S_{z}^{(r,t)}\left(z\right)=\left(c/8\pi\right)\intop_{0}^{D/2}dx\thinspace E_{x}^{(r,t)}\left(x,z\right)\overline{H_{y}^{(r,t)}\left(x,z\right)}.\label{eq:Srt}
\end{eqnarray}
Electromagnetic fields of the reflected wave can be obtained from
the second column of the 2$\times$2 matrix in Eq.\,\eqref{eq:HEv-vacuum}, and after substituting
them into Eq.\,(\ref{eq:Srt}), the respective component of Poynting
vector has the form 
\begin{eqnarray}
S_{z}^{(r)}=\frac{cD}{16\pi}\mathrm{Re}\left\{ \sum_{m=0}^{\infty}\left(\hat{F}_{m}\right)_{22}\frac{1+\delta_{m,0}}{2}\left|H_{y||m}^{(r)}\right|^{2}\right\} .\label{eq:Szr-vacuum}
\end{eqnarray}
In a similar manner for the transmitted wave {[}whose electromagnetic
field can be obtained from the first column of the 2$\times$2 matrix in Eq.\,(\ref{eq:HEs-substrate}){]},
we have
\begin{eqnarray}
S_{z}^{(t)}=\frac{cD}{16\pi}\mathrm{Re}\left\{ \sum_{m=0}^{\infty}\left(\hat{F}_{m}\right)_{21}\frac{1+\delta_{m,0}}{2}\left|H_{y||m}^{(t)}\right|^{2}\right\} .\label{eq:Szt-vacuum}
\end{eqnarray}
Notice, that $S_{z}^{(r,t)}$ in vacuum are independent of distances
owing to the absence of losses. Taking into account the Poynting vector's
$z$-component of incident wave $S_{z}^{(i)}=cD\left|H_{y}^{(i)}\right|^{2}/(16\pi)$
{[}whose electromagnetic field can be taken from the first column
of the 2$\times$2 matrix in Eq.\,(\ref{eq:HEv-vacuum}){]}, we define the integral reflectance $R$
and transmittance $T$ of the structure as relations 
\begin{eqnarray}
R=-\frac{S_{z}^{(r)}}{S_{z}^{(i)}}=-\frac{\mathrm{Re}\left\{ \sum_{m=0}^{\infty}\left(\hat{F}_{m}\right)_{22}\frac{1+\delta_{m,0}}{2}\left|H_{y||m}^{(r)}\right|^{2}\right\} }{\left|H_{y}^{(i)}\right|^{2}},\label{eq:reflectance}\\
T=\frac{S_{z}^{(t)}}{S_{z}^{(i)}}=\frac{\mathrm{Re}\left\{ \sum_{m=0}^{\infty}\left(\hat{F}_{m}\right)_{21}\frac{1+\delta_{m,0}}{2}\left|H_{y||m}^{(t)}\right|^{2}\right\} }{\left|H_{y}^{(i)}\right|^{2}}.\label{eq:transmittance}
\end{eqnarray}
The minus sign in Eq.\,(\ref{eq:reflectance}) accounts for the
negative direction of the reflected wave propagation.

\section{Eigenmodes of surface phonon-plasmon-polaritons}

The dispersion relation for eigenmodes of the equivalent structure
with continuous graphene, shown in Fig.\,\ref{fig:geometry}(b) can
formally be obtained from Eqs.\,\eqref{eq:bc-H-res} and \eqref{eq:bc-E-res}
by formal substitution $2\pi m/D=k_{x}$ and Ohm's law for continuous
graphene $j_{x||m}=\sigma_{g}\left(\omega\right)\left(\hat{F}_{m}^{(+,tot)}\right)_{21}H_{y||m}^{(t)}$.
For the case without external wave $H_{y}^{(i)}\equiv0$ (condition
for the eigenmodes), the dispersion relations will have the form
\begin{eqnarray}
1+\frac{4\pi}{cg_{m}}\sigma_{g}\left(\omega\right)=0.\label{eq:disp-rel}
\end{eqnarray}

\begin{figure*}[t]
	\centering\includegraphics[width=17cm]{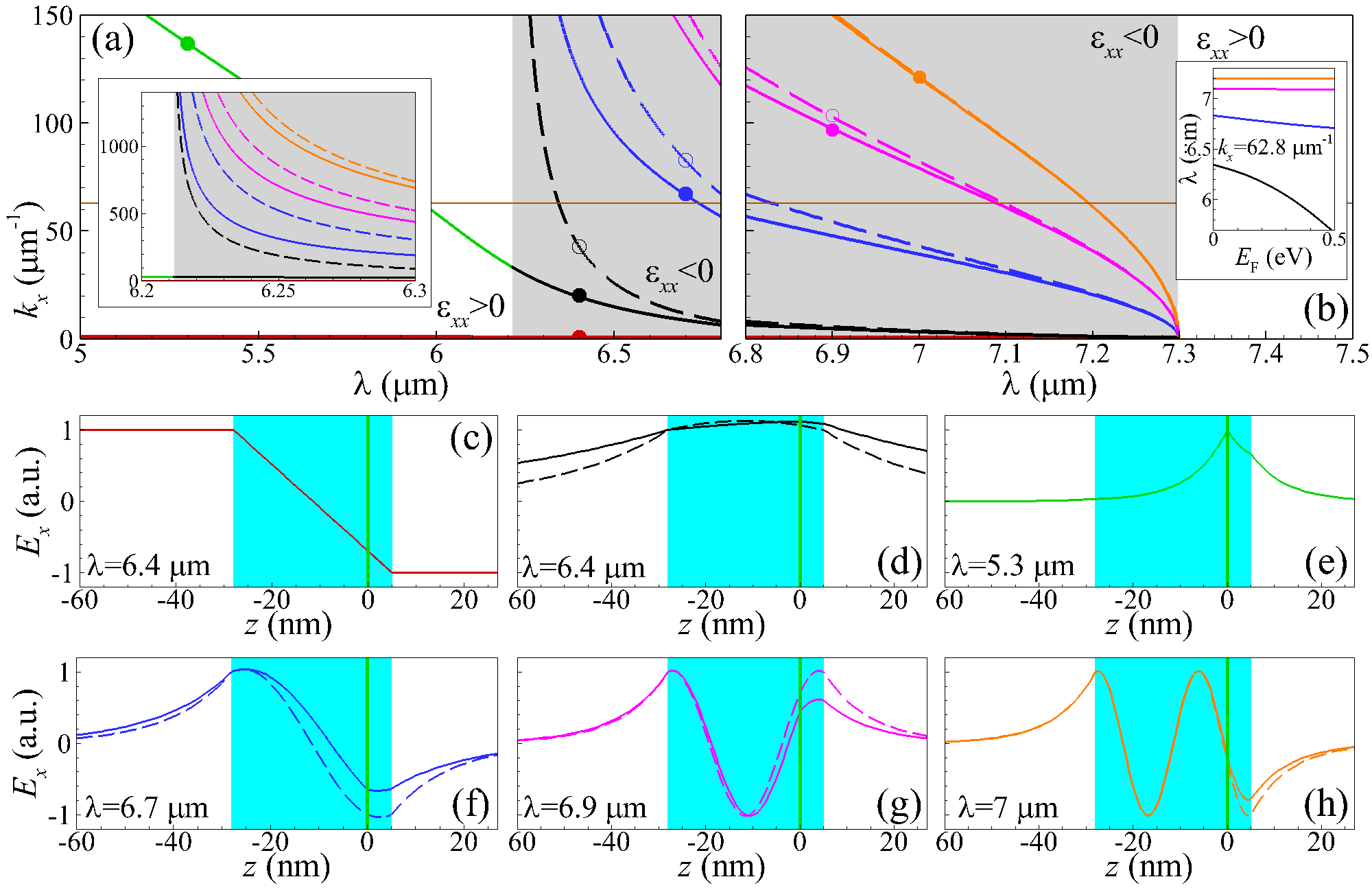}
	
	\caption{(a)-(b) Spectrum of the phonon-plasmon polaritons (five lower modes) in the structure with continuous graphene [shown in Fig.\,\ref{fig:geometry}(b)]
		in the vicinity of hBN upper restrahlen band for undoped graphene
		(dashed lines) and doped graphene (solid lines). Horizontal
		brown solid line correspond to the wavevector $k_{x}=62.8\thinspace\mu\mathrm{m}^{-1}$. In panel (a) inset shows zoom of dispersion curves near asymptotics,
		while inset in panel (b) demonstrates the dependence of eigenmode
		free-space wavelength upon Fermi energy for fixed value of in-plane
		wavevector $k_{x}$; (c)--(h) Spatial profiles of the electric field's
		$x$-component $E_{x}$ of first {[}panel (c){]}, second {[}panels
		(d) and (e){]}, third {[}panel (f){]}, forth {[}panel (g){]} and fifth
		{[}panel (h){]} eigenmodes for undoped (dashed lines) and doped (solid
		lines) graphene. Parameters (free-space wavelength $\lambda$ and
		in-plane wavevector $k_{x}$) of eigenmodes, depicted in panels (c)--(h)
		are shown in panel (a) and (b) by filled (empty) circles for doped
		(undoped) graphene. Areas, occupied by h-BN are shaded in panels (c)--(h)
		by blue color, while position of graphene is depicted by green vertical
		line. In all panels other parameters are: $d_{t}=28\thinspace$nm,
		$d_{b}=5\thinspace$nm, and the Fermi energy of doped graphene $E_{F}=0.35\thinspace$eV. }
	
	\label{fig:eigenfreq_restr_band_up}
\end{figure*}

The spectrum of the equivalent structure {[}Fig.\,(\ref{fig:geometry})(b){]}
can be obtained as solution of the dispersion relation (\ref{eq:disp-rel}),
and in the vicinity of the upper RB consists of infinite number of
modes {[}first five of which are shown in Figs.\,\ref{fig:eigenfreq_restr_band_up}(a),
\ref{fig:eigenfreq_restr_band_up}(b){]}, which start at the lower\footnote{Lower and upper edges are characterized in terms of frequency, hence,
	the free-space wavelength of upper edge is shorter than that of lower
	edge.} edge of RB $\lambda\approx7.3\,\mu$m (where $\varepsilon_{xx}=0$).
In the case of the undoped graphene all the spectrum {[}dashed lines{]} except the first mode [red dashed line]
lies in the frequency domain $6.22\,\mu\mathrm{m}\lesssim\lambda\lesssim7.3\,\mu$m,
where $\varepsilon_{xx}<0$, and asymptotically approaches the RB
upper edge {[}see inset in Fig.\,\ref{fig:eigenfreq_restr_band_up}(a){]}.
The first mode with lowest $k_{x}$ {[}red line in Figs.\ref{fig:eigenfreq_restr_band_up}(a)
and \ref{fig:eigenfreq_restr_band_up}(b){]} lies nearby the light
line in vacuum $k_{x}=\omega/c$. As a result, it is characterized
by weak decay of its electric field inside the vacuum and is antisymmetric
inside hBN-graphene slab {[}see spatial profile in Fig.\,\ref{fig:eigenfreq_restr_band_up}(c){]}.
Other modes differ from each other by the number of variations of
sign of electric field across the hBN layer {[}as shown in Figs.\,\ref{fig:eigenfreq_restr_band_up}(c)--\ref{fig:eigenfreq_restr_band_up}(h){]},
which is typical for dielectric waveguide modes (further in the paper
such kind of modes will be referred as waveguide-like). Thus, the
second mode {[}black dashed lines in Figs.\,\ref{fig:eigenfreq_restr_band_up}(a)--\ref{fig:eigenfreq_restr_band_up}(b){]}
is characterized by absence of sign variation, while third (blue line),
fourth (pink line) and fifth (orange line) modes exhibit one, two,
and three sign variation, respectively {[}see Figs.\,\ref{fig:eigenfreq_restr_band_up}(d),
\ref{fig:eigenfreq_restr_band_up}(f), \ref{fig:eigenfreq_restr_band_up}(g),
and \ref{fig:eigenfreq_restr_band_up}(h){]}. Doping of graphene results
in the lowering of free-space wavelengths $\lambda$ of all modes {[}see inset in Fig.\,\ref{fig:eigenfreq_restr_band_up}(b){]}.
At the same time, only third and higher modes {[}solid blue, pink, and orange lines in Figs.\,\ref{fig:eigenfreq_restr_band_up}(a)--\ref{fig:eigenfreq_restr_band_up}(b)
{]} asymptotically approaches the upper edge of RB, while the second
mode (solid black line) at RB's upper edge transforms into the plasmon-like
mode (green line), which exists beyond the RB and is absent in the
case of undoped graphene. The plasmon-like nature of this mode is
defined by the fact that its electromagnetic field {[}see Fig.\,\ref{fig:eigenfreq_restr_band_up}(e){]}
is localized nearby the graphene layer, while waveguide-like modes
are localized inside the hBN slab {[}solid lines in Figs.\,\ref{fig:eigenfreq_restr_band_up}(d),
\ref{fig:eigenfreq_restr_band_up}(f), \ref{fig:eigenfreq_restr_band_up}(g),
and \ref{fig:eigenfreq_restr_band_up}(h){]}. Notice, that graphene's
doping slightly perturbs the electromagnetic field spatial distributions
{[}compare dashed and solid lines in Figs.\,\ref{fig:eigenfreq_restr_band_up}(d),
\ref{fig:eigenfreq_restr_band_up}(f), \ref{fig:eigenfreq_restr_band_up}(g),
and \ref{fig:eigenfreq_restr_band_up}(h){]}, while numbers of sign
variations are kept unchanged. 

\begin{figure*}[t]
	\centering\includegraphics[width=17cm]{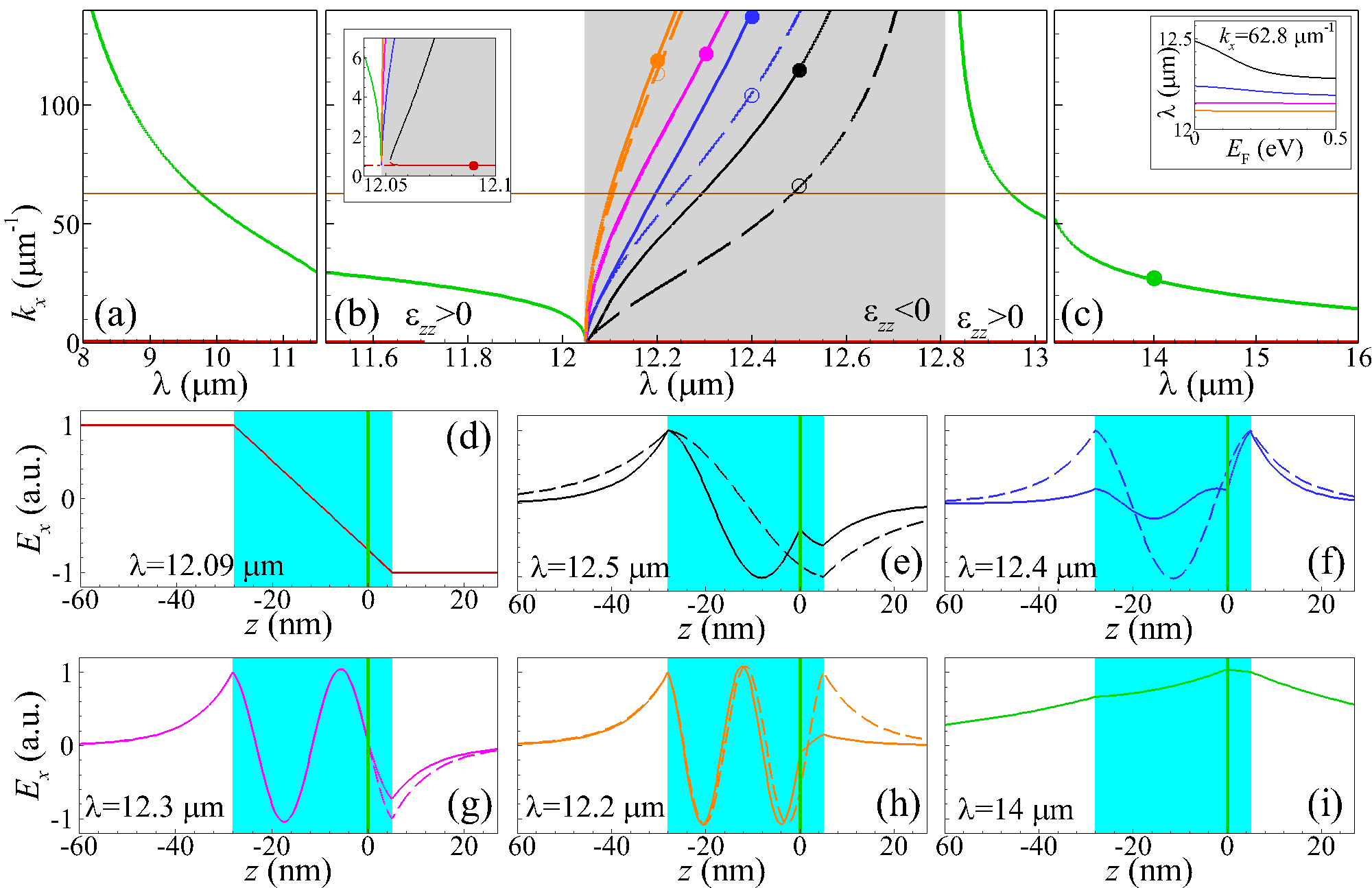}
	
	\caption{(a)-(c) Spectrum of the phonon-plasmon polaritons (five lower modes)  in the structure with continuous graphene [shown in Fig.\,\ref{fig:geometry}(b)]
		in the vicinity of hBN lower restrahlen band for undoped graphene
		(dashed lines) and doped (solid lines) graphene (solid lines). In
		panel (b) inset shows zoom of dispersion curves near the bifurcation
		point between first and second modes, while inset in panel (c) demonstrates
		the dependence of eigenmode free-space wavelength upon Fermi energy
		for fixed value of in-plane wavevector $k_{x}$; (c)--(h) Spatial
		profiles of the electric field's $x$-component $E_{x}$ of first
		{[}panel (d){]}, second {[}panel (e){]}, third {[}panel (f){]}, forth
		{[}panel (g){]}, and fifth {[}panel (h){]} eigenmodes and as well
		as eigenmode inside the gap {[}panel (i){]} for undoped (dashed lines)
		and doped (solid lines) graphene. Parameters (free-space wavelength
		$\lambda$ and in-plane wavevector $k_{x}$) of eigenmodes, depicted
		in panels (d)--(i) are shown in panel (a) and (b) by filled (empty)
		circles for doped (undoped) graphene. Other parameters as well as
		meaning of horizontal lines in panels (a)--(c) and of shaded areas
		and vertical lines in panels (d)--(i) are the same as those in Fig.\,\ref{fig:eigenfreq_restr_band_up}.}
	
	\label{fig:eigenfreq_restr_band_low}
\end{figure*}

Another plasmonic-like mode exist in the gap between upper and lower
RBs {[}green line in Figs.\,\ref{fig:eigenfreq_restr_band_low}(a) and \ref{fig:eigenfreq_restr_band_low}(b){]},
which starts at the upper edge of the lower RB. At the same time,
third and higher waveguide modes {[}blue, pink, and orange lines in
Fig.\,\ref{fig:eigenfreq_restr_band_low}(b){]} also have origin
at the upper edge of the lower RB, while the first and second waveguide
modes (depicted by red and black lines, respectively) have common
bifurcation point {[}see inset in Fig.\,\ref{fig:eigenfreq_restr_band_low}(b){]}.
All the waveguide modes (except first one) approach asymptotically
lower edge of the lower RB, while first one exists inside the lower
gap both in the case of doped and undoped graphene. At the same time,
for the case of doped graphene lower gap contains also a plasmon-like
mode {[}depicted by green solid line in Figs.\,\ref{fig:eigenfreq_restr_band_low}(b)
and \ref{fig:eigenfreq_restr_band_low}(c), and which field is shown
in Fig.\,\ref{fig:eigenfreq_restr_band_low}(i){]}. Similar to the
case of upper RB graphene doping lowers the free-space wavelengths $\lambda$ of all modes
{[}see inset in Fig.\,\ref{fig:eigenfreq_restr_band_low}(c){]},
and perturbs the field spatial distribution {[}Figs.\,\ref{fig:eigenfreq_restr_band_low}(d)--\ref{fig:eigenfreq_restr_band_low}(h){]}.
Along with this,unlike the previous case (upper RB), in the
case of lower RB there is no mode with absence of sign variation.

\section{Absorbance spectrum of graphene nanoribbon periodic array}

\begin{figure*}[t]
	\centering\includegraphics[width=17cm]{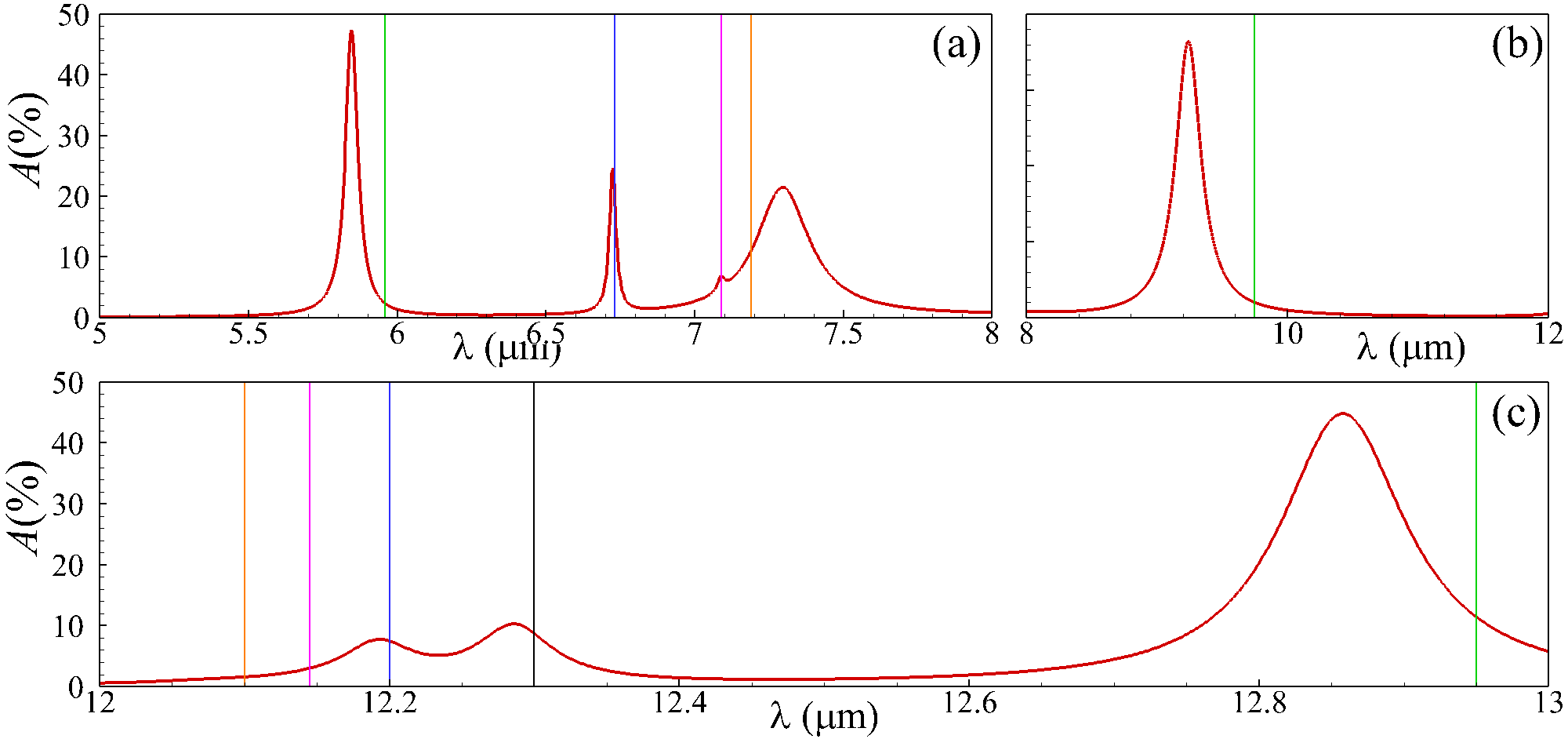}
	
	\caption{Dependence of plane-wave absorbance $A$ upon free-space wavelength $\lambda$
		of the periodical structure without analyte layer [depicted in Fig.\,\ref{fig:geometry}(a)]. The parameters 
		of the structure are: $D=100\thinspace$nm, $W=50\thinspace$nm, $E_{F}=0.7\thinspace$eV, $d_{a}=0$, $d_{t}=28\,$nm,
		$d_{b}=5\,$nm. Vertical
		lines depict free-space wavelengths of eigenmodes of equivalent structure [depicted in Fig.\,\ref{fig:geometry}(b)] for fixed wavevector $k_{x}=2\pi/D$, namely second (black
		lines), third (blue lines), forth (pink lines), and fifth (orange lines)  inside
		the restranlen bands 
		and eigenmode outside restrahlen bands (green lines). }
	
	\label{fig:A(lambda)}
\end{figure*}

All the eigenmodes (both waveguide and plasmonic), depicted in 
Figs.\,\ref{fig:eigenfreq_restr_band_up}(c)--\ref{fig:eigenfreq_restr_band_up}(h)
and \ref{fig:eigenfreq_restr_band_low}(d)--\ref{fig:eigenfreq_restr_band_low}(i) are evanescent waves inside the spatial domains occupied by vacuum.
This fact means the impossibility to couple these modes directly to an
external propagating wave. However, they can be coupled to the diffracted harmonics
of a periodic grating. In this case the in-plane component of the incident
wave will be effectively elongated by an integer number of the reciprocal
grating vectors. For example, if $D$ is the period of the grating,
then normally incident external wave can excite those modes in the
spectrum, which wavevectors are equal to $k_{x}=2\pi m/D$ (here $m$
is an integer number). For the particular value of the period $D=100\,$nm
the free-space wavelengths, at which mode excitation is possible,
can be estimated as crossing points between the spectrum {[}shown
in Figs.\,\ref{fig:eigenfreq_restr_band_up}(a), \ref{fig:eigenfreq_restr_band_up}(b),
\ref{fig:eigenfreq_restr_band_low}(a)--\ref{fig:eigenfreq_restr_band_low}(c){]}
and horizontal solid brown lines (which correspond to the wavevector
$k_{x}=2\pi/D$ of first harmonic with $m=1$). The values of these
predicted free-space wavelengths are depicted by vertical lines in
Fig.\,\ref{fig:A(lambda)}, while the calculated absorbance is shown
by the solid red line. This absorbance spectrum is calculated for the
Fermi energy of graphene strips $E_{F}=0.7\thinspace$eV in order
to have average Fermi energy on period $E_{F}(W/D)$ equal to that,
used in Figs.\,\ref{fig:eigenfreq_restr_band_up} and \ref{fig:eigenfreq_restr_band_low}.
As it is seen, the absorbance spectrum contains various peaks, which
correspond to different eigenmodes excitation (for comparison, in the case of graphene without hBN the absorbance spectrum would contain only one peak, see Ref.\cite{gr-nanoribbons-analytical-Goncalves2016-prb}). It is interesting,
that positions of peaks coincide well with the eigenmode frequencies
for the case of waveguide-like modes {[}blue and pink solid lines
in Fig.\,\ref{fig:A(lambda)}(a) as well blue and black lines in
Fig.\,\ref{fig:A(lambda)}(c){]}, but slightly differs from the predicted
ones in the case of plasmonic-like modes. The reason for such a behaviour
is that waveguide-like modes are localized inside the whole hBN slab,
thus influence of graphene patterning on mode properties being relatively
small. At the same time, the plasmon-like modes are localized in the
vicinity of graphene, where graphene pattering produces strong influence
on the mode properties. An analysis of field distributions and their correspondence to eigenmode excitation is described in Supplemental document.

The position of absorption peaks, which correspond both to the excitation
of waveguide-like and plasmon-like modes depend strongly upon the
position of graphene inside the hBN slab, as it can be seen from Fig.\,\ref{fig:A(lambda,rel)}.
Thus, when graphene is relocated from the edge of the slab into its
center, the position of the absorbance peaks are red-shifted. 

\begin{figure}[t]
	\centering\includegraphics[width=8.5cm]{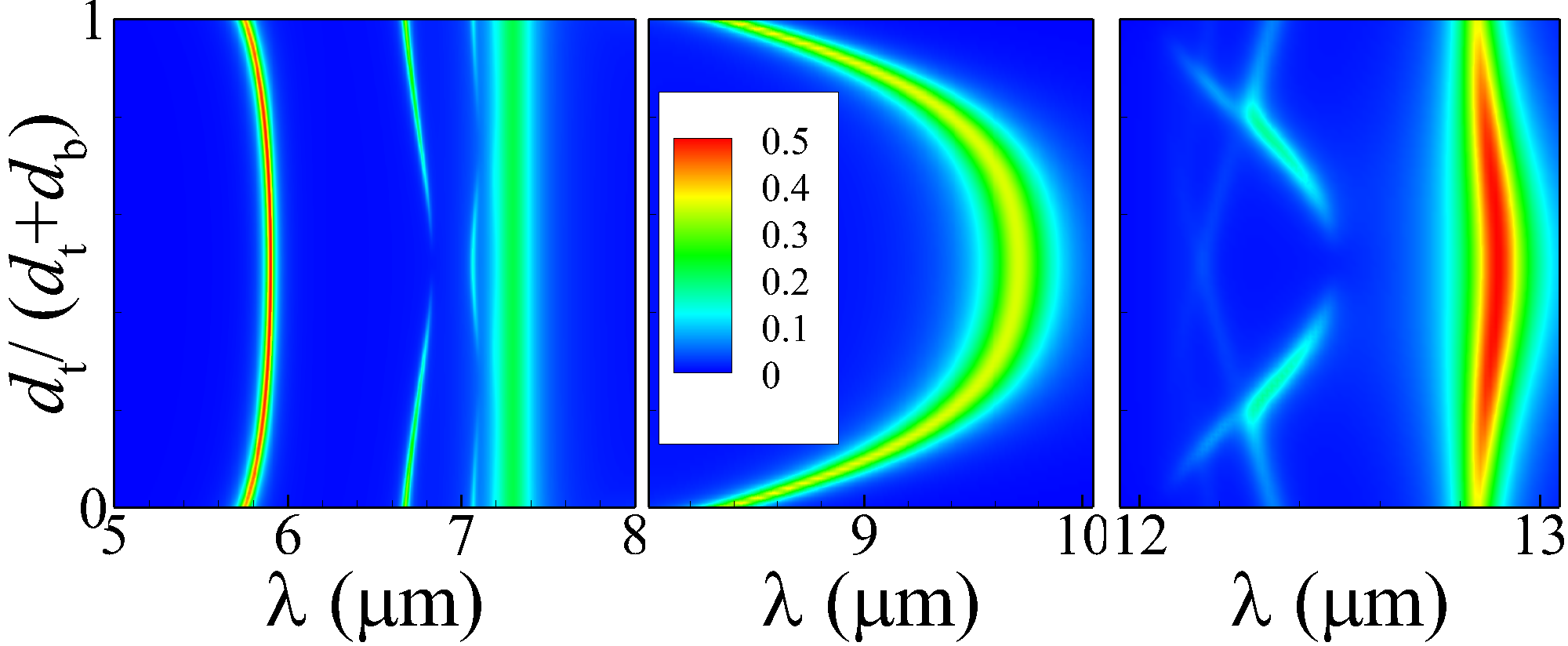}
	
	\caption{Plane-wave absorbance $A$ versus free-space wavelength $\lambda$
		and relation between the thicknesses of top and bottom hBN layers
		$d_{t}/\left(d_{b}+d_{t}\right)$ for the structure without analyte
		layer $d_{\mathrm{anal}}=0$. The total thickness of structure is
		kept fixed $d_{t}+d_{b}=33\thinspace$nm. Other parameters are the
		same as those in Fig.\,\ref{fig:A(lambda)}. }
	
	\label{fig:A(lambda,rel)}
\end{figure}

In order to clarify the possibility of using this structure for 
sensing, we demonstrate in Fig.\,\ref{fig:A(lambda,Danal)},
that adding of the analyte layer on top of structure shifts positions
of absorbance peaks. For the particular choice of parameters in Fig.\,\ref{fig:A(lambda,Danal)},
highest sensitivity to the variation of analyte layer thickness is
observed for absorbance peaks nearby $\lambda\sim6.2\thinspace\mu$m,
$\lambda\sim9.6\thinspace\mu$m, $\lambda\sim12.35\thinspace\mu$m,
and $\lambda\sim12.9\thinspace\mu$m. These peaks correspond to the
excitation of plasmon-like mode for wavelength below upper RB {[}green
line in Fig.\,\ref{fig:eigenfreq_restr_band_up}(a){]}, plasmon-like
mode between RBs {[}green lines in Figs.\,\ref{fig:eigenfreq_restr_band_low}(a)
and \ref{fig:eigenfreq_restr_band_low}(b){]}, second waveguide-like
mode in lower RB {[}black line in Fig.\,\ref{fig:eigenfreq_restr_band_low}(b){]}
and plasmon-like mode below lower RB {[}green line in Figs.\,\ref{fig:eigenfreq_restr_band_low}(b)
and \ref{fig:eigenfreq_restr_band_low}(c){]}, respectively. For these
peaks the significant shift takes place already at the analyte thicknesses
$d_{\mathrm{a}}\sim20\thinspace$nm, while further increase of
analyte thickness does not influence position of these peaks, owing
to the rapid decay of field intensity away from interface between
top hBN layer and analyte. At the same time the absorbance peaks nearby
$\lambda\sim6.8\thinspace\mu$m, which corresponds to the excitation
of waveguide-like mode in upper RB {[}blue line in Figs.\,\ref{fig:eigenfreq_restr_band_up}(a)
and \ref{fig:eigenfreq_restr_band_up}(b){]} is much less sensitive
to the presence of analyte layer, while the peak nearby $\lambda\sim7.3\thinspace\mu$m
(mode bunching nearby edge of RB) is almost insensitive. Notice, with
an increase of analyte layer thickness two waveguide-like modes {[}corresponding
to the black and blue lines in Fig.\ref{fig:eigenfreq_restr_band_low}(b){]}
approximate to each other. The presence of highly analyte-sensitive and virtually analyte-insensitive absorption bands at the same spectral range can be beneficial for chemical/biosensing applications, as the insensitive bands can serve as a reference, as well as allow for reducing cross-sensitivity with, e.g., temperature variations.
\begin{figure}[t]
	\centering\includegraphics[width=8.5cm]{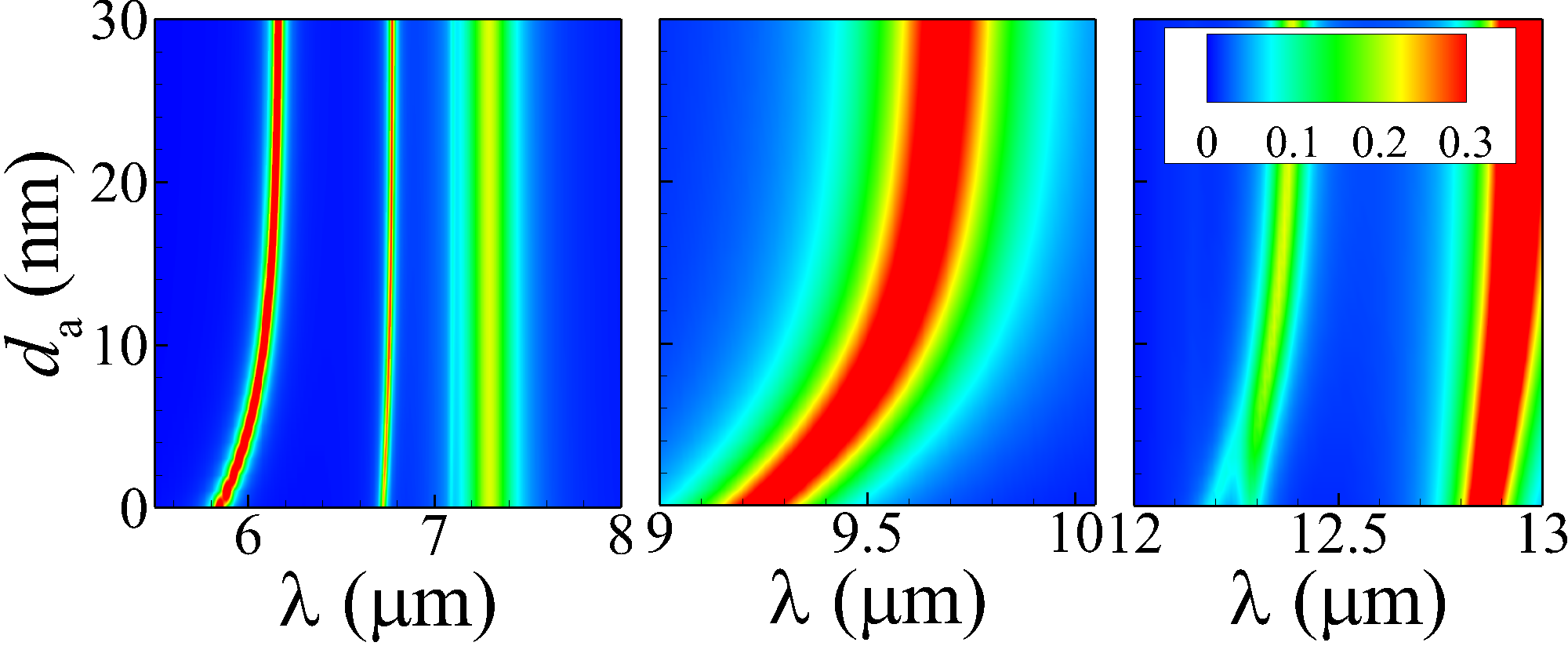}
	
	\caption{Absorbance (depicted by color map) versus free-space wavelength $\lambda$
		and thickness of analyte layer $d_{\mathrm{a}}$ for the case,
		where analyte refractive index $n_{\mathrm{a}}=2$. Other parameters
		are the same, as those in Fig.\,\ref{fig:A(lambda)}.}
	
	\label{fig:A(lambda,Danal)}
\end{figure}

In order to optimise the structure, in Fig.\,\ref{fig:A(lambda,rel)_anal}
we depict the dependence of absorbance upon the relative position
of graphene inside the hBN layer. From this figure it is clear, that
maximal shift of the plasmon-like mode below RBs (nearby $\lambda\sim13\thinspace\mu$m)
takes place, when graphene is arranged in the center of the structure
($d_{t}\approx d_{b}$). At the same time, for plasmon-like modes
above RBs (nearby $\lambda\sim6\thinspace\mu$m) and between RBs (nearby
$8\mu$m$\lesssim\lambda\lesssim 10\thinspace\mu$m) relative shifts of absorbance
peaks (if compared to one without analyte layer, which absorbance
maxima are shown by white lines) are maximal, when graphene is close
to the analyte layer. Thus, when top hBN layer is absent, $d_t=0$, the maximal shift of resonance at $n_{\mathrm{a}}=1$ and $n_{\mathrm{a}}=2$ is $\Delta\lambda\approx9.37\,\mu$m$-8.37\,\mu$m$=1\mu$ m, or $\Delta\omega\approx130$cm$^{-1}$, which is compatible to experimentally observed in Ref.\cite{spp-gr-nanoribbons-sensor-Rodrigo2015-science}.
\begin{figure}[t]
	\centering\includegraphics[width=8.5cm]{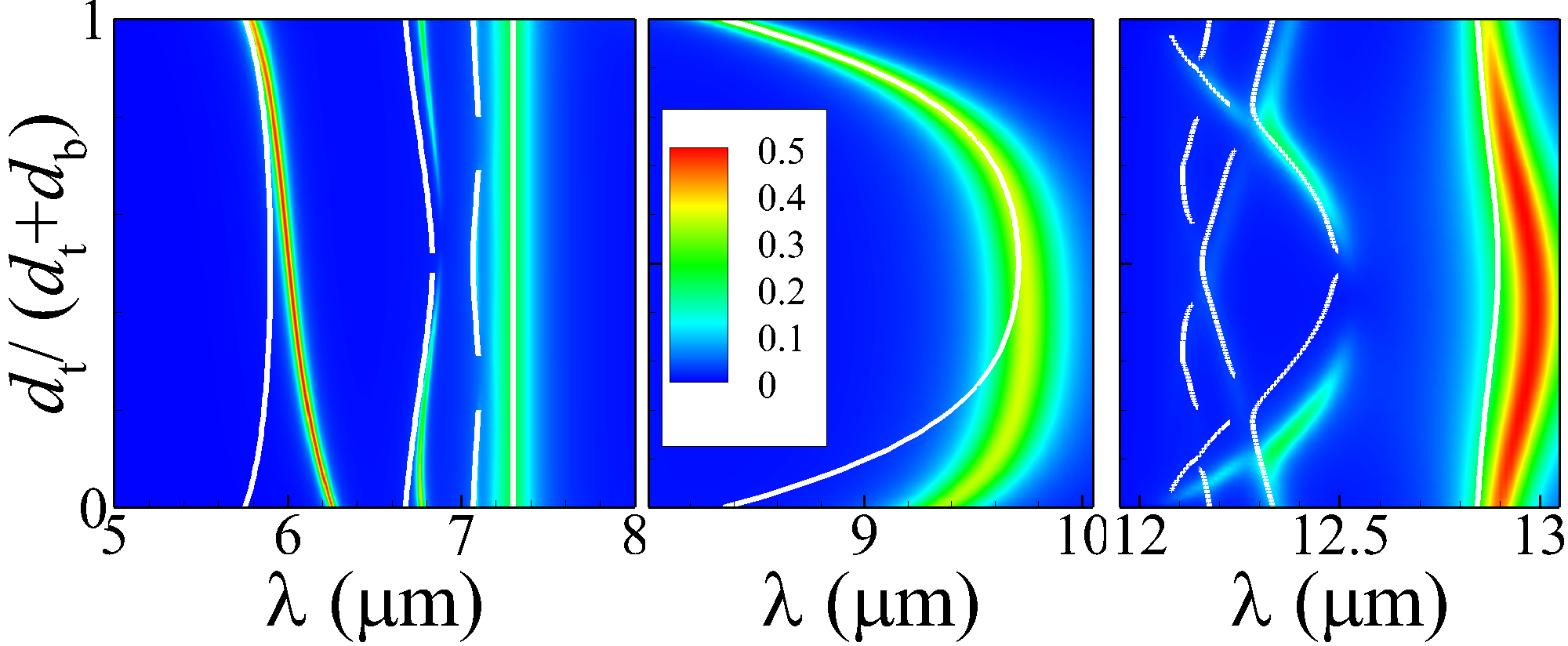}
	
	\caption{Plane-wave absorbance $A$ versus free-space wavelength $\lambda$
		and relation between the thicknesses of top and bottom hBN layers
		$d_{t}/\left(d_{b}+d_{t}\right)$ for the structure with analyte layer
		of thickness $d_{\mathrm{a}}=20\thinspace$nm and refractive index
		$n_{\mathrm{a}}=2$. The total thickness of structure is kept fixed
		$d_{t}+d_{b}=33\thinspace$nm. Other parameters are the same as those
		in Fig.\,\ref{fig:A(lambda)}. The position of absorbance peak maxima
		of structure without analyte layer are depicted by white solid lines. }
	
	\label{fig:A(lambda,rel)_anal}
\end{figure}
In more details this phenomenon is shown in middle column of Fig.\,\ref{fig:A(lambda,Nanal)},
where for $d_{t}=30\thinspace$nm {[}Fig.\,\ref{fig:A(lambda,Nanal)}(b1){]}
the shift of resonance at $n_{\mathrm{a}}=1$ and $n_{\mathrm{a}}=2$
is indistinguishable in the scale of panel, while for $d_{t}=3\thinspace$nm
and the same analyte layer thickness {[}Figs.\,\ref{fig:A(lambda,Nanal)}(b2){]}
the sensitivity\cite{sensor-Wen2019-optCom} $S=\Delta\lambda/\Delta n_{\mathrm{a}}\approx560\thinspace\mathrm{nm/RIU}$ with figure-of-merit ${\rm FoM}=S/\Delta \lambda_{0.5}\approx 2.6\,\mathrm{RIU}^{-1}$, with $\Delta \lambda_{0.5}$ being the width of the resonance at half-maximum\cite{spp-FoM-Meng2017-sensors}. This value of sensitivity exceeds the theoretical estimation, given in Ref.\cite{spp-gr-nanoribbons-sensor-Vasic2013-jap}, while FoM is compatible with Ref.\cite{spp-gr-nanoribbons-sensor-Vasic2013-jap}.
For the thinner analyte layer $d_{\mathrm{a}}=5\thinspace$nm {[}Fig.\,\ref{fig:A(lambda,Nanal)}(b3){]}
the sensitivity is diminished considerably $S\approx250\thinspace\mathrm{nm/RIU}$.
The same situation takes place for plasmon-like mode above RB: when
graphene is arranged nearby the hBN layer's edge, opposite to analyte
layer {[}Fig.\,\ref{fig:A(lambda,Nanal)}(a1){]}, the sensitivity
of this mode is $S\approx30\thinspace\mathrm{nm/RIU}$. Nevertheless,
when graphene is close to the analyte layer {[}Fig.\,\ref{fig:A(lambda,Nanal)}(a2){]}
the sensitivity becomes considerably larger $S\approx370\thinspace\mathrm{nm/RIU}$ [with ${\rm FoM}\approx 7.75\,\mathrm{RIU}^{-1}$ which is considerably higher if compared to that in Fig.\,\ref{fig:A(lambda,Nanal)}(b2) owing to narrower resonance]
for the same analyte layer thickness and $S\approx200\thinspace\mathrm{nm/RIU}$
for thinner analyte layer {[}see Fig.\,\ref{fig:A(lambda,Nanal)}(a3){]}.
At the same time, the sensitivity of third waveguide-like mode in
lower RB, located nearby $\lambda\sim6.7\thinspace\mu$m, is almost
equal $S\approx60\thinspace\mathrm{nm/RIU}$ both for the case $d_{t}=30\thinspace$nm,
$d_{b}=3$nm {[}Fig.\,\ref{fig:A(lambda,Nanal)}(a1){]} and for the
case $d_{t}=3\thinspace$nm, $d_{b}=30$nm {[}Fig.\,\ref{fig:A(lambda,Nanal)}(a2){]},
owing the the field distribution inside all hBN slab. In the upper
RB an interesting phenomenon takes place: when graphene layer is located
nearby analyte {[}Figs.\,\ref{fig:A(lambda,Nanal)}(c2) and \ref{fig:A(lambda,Nanal)}(c3){]}
the second waveguide-like mode, located nearby $\lambda\sim12.3\thinspace\mu$m
is almost insensitive to the presence of analyte layer, while third
waveguide-like mode (nearby $\lambda\sim12.1\thinspace\mu$m) exhibits
the sensitivities $S\approx110\thinspace\mathrm{nm/RIU}$ and $S\approx40\thinspace\mathrm{nm/RIU}$
for $d_{\mathrm{a}}=20\thinspace$nm and $d_{\mathrm{a}}=5\thinspace$nm,
respectively. At the same time, when graphene layer is arranged at
the edge of hBN slab, opposite to analyte layer {[}Figs.\,\ref{fig:A(lambda,Nanal)}(c1){]}
the situation is opposite: here second-waveguide-like mode is almost
insensitive to the presence of analyte layer, while third one exhibits
the sensitivity $S\approx50\thinspace\mathrm{nm/RIU}$. 
\begin{figure*}[t]
	\centering\includegraphics[width=17cm]{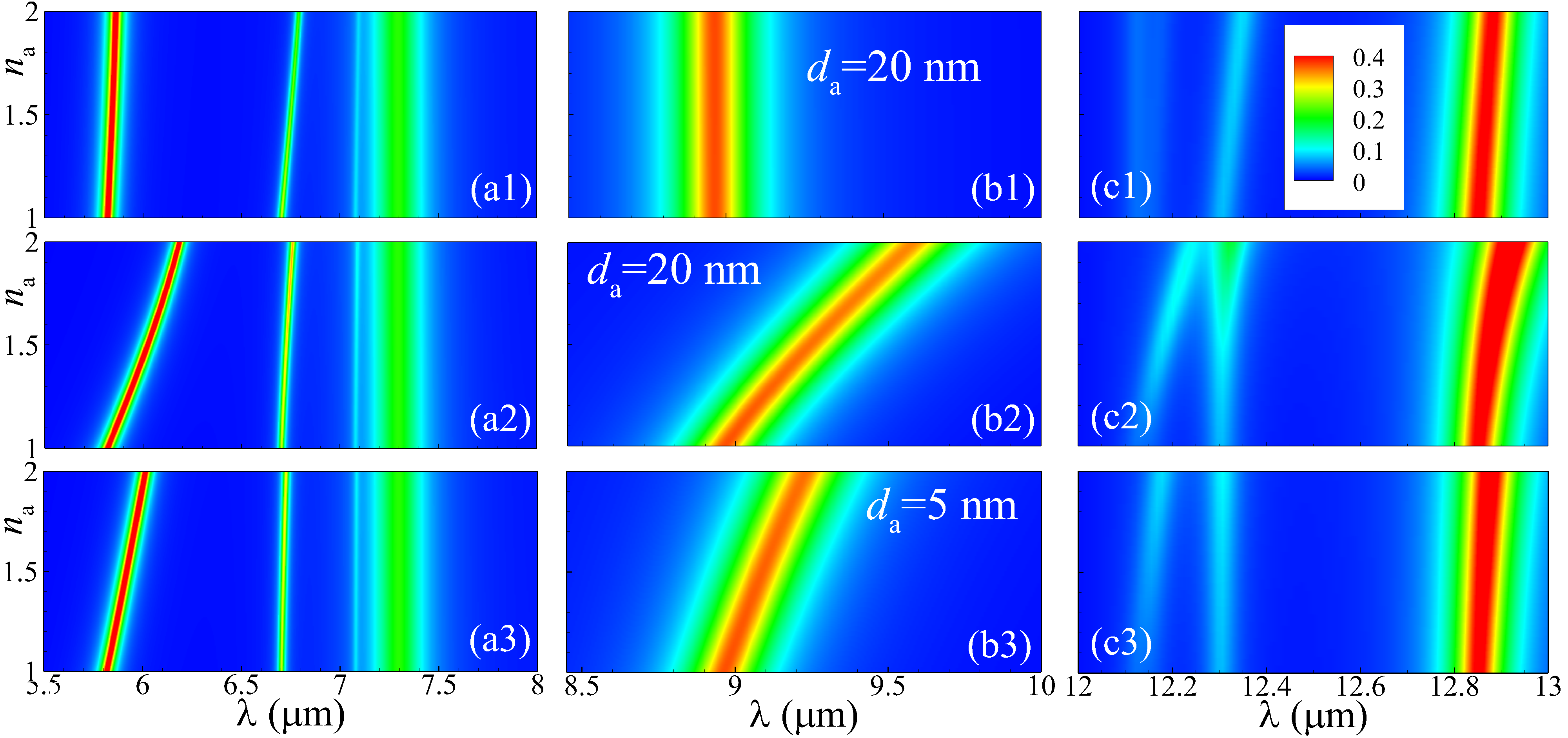}
	
	\caption{Absorbance (depicted by color map) versus free-space wavelength $\lambda$
		and analyte refractive index $n_{\mathrm{a}}$ for two values of
		analyte layer thickness $d_{\mathrm{a}}=20\thinspace$nm {[}top
		and middle rows, panels (a1)--(c1), (a2)--(c2){]}, and $d_{\mathrm{a}}=5\thinspace$nm
		{[}bottom row, panels (a3)--(c3){]} and hBN layers thicknesses $d_{t}=30\thinspace$nm,
		$d_{b}=3\thinspace$nm {[}top row, panels (a1)--(c1){]} and $d_{t}=3\thinspace$nm,
		$d_{b}=30\thinspace$nm {[}middle and bottom rows, panels (a2)--(c2),
		(a3)--(c3){]}. Other parameters are the same, as those in Fig.\,\ref{fig:A(lambda)}.}
	
	\label{fig:A(lambda,Nanal)}
\end{figure*}

\section{Conclusions}
In the present paper we investigated theoretically the coupling of an external plane electromagnetic wave to diffraction grating, consisting of a periodic array of graphene nanoribbons, cladded by two layers of hBN. We demonstrate, that the absorbance spectrum of this structure exhibits a series of peaks, related to the excitation of eigenmodes. We also showed, that this structure can be potentially used in the sensing applications. In detail, deposition of an analyte dielectric layer of relatively small thickness $\sim 20\,$nm on top of this structure results in the considerable shift of several absorbance peak positions. The peak, related to the excitation of the plasmon-like eigenmode between RBs, demosntrates the maximal shift $\sim 1000\,$nm/RIU  in the case, when graphene nanoribbon array is located in the vicinity of analyte layer. Further increase of sensitivity can be achieved by the patterning of graphene in more complicated manner, like flower-shape\cite{spp-gr-sensor-flower-Razani2022-optcomm}, split-ring resonators \cite{spp-gr-sensor-split-ring-Chen2020-mre}, crosses \cite{spp-gr-sensor-cross-Liu2022-nrl,spp-gr-sensor-cross-BarzegarParizi2022-oqe} among others. Another potential application of these graphene-hBN hybrid structures is to be building blocks of a new class of photodetectors \cite{photodetector-gr-hBN-Castilla2020-natcom,photodetector-gr-hBN-Woessner2017-2Dmat}, where these structures serve simultaniously as active media, sustaining phonon-plasmon polaritons and as key-element of the detector.

\section*{Acknowledgments} 

C.~J.~S.~dM. and D.~A.~B. acknowledge financial support by FAPESP (grant no. 2018/07276-5), FINEP (grant no. 01.22.0208.00), the Brazilian Nanocarbon Institute of Science and Technology (INCT/Nanocarbon), and CAPES-PrInt (grant nos. 88887.310281/2018-00 and 88881.310294/2018-01). Y.~V.~B. and N.~M.~R.~P. acknowledge support by the Portuguese Foundation for Science and Technology (FCT) in the framework of the Strategic Funding UIDB/04650/2020, COMPETE 2020, PORTUGAL 2020, FEDER. N.~M.~R.~P. also acknowledges support by FCT through projects PTDC/FIS-MAC/2045/2021 (10.54499/PTDC/FIS-MAC/2045/2021) and EXPL/FIS-MAC/0953/ 2021, as well as the Independent Research Fund Denmark (grant no. 2032-00045B) and the Danish National Research Foundation (Project No.~DNRF165).

\section*{Disclosures} 
The authors declare no conflicts of interest.

\bibliographystyle{unsrtnat}
\bibliography{hBN_graphene_bib}

\section{Hyperbolic phonon-plasmon polaritons in a hBN-graphene van der Waals structure: supplemental document}

	\subsection{Spatial distributions of electric field of diffracted wave}

\begin{figure}[htbp]
		\centering
		%\fbox{\includegraphics[width=.6\linewidth]{sample}}
		\fbox{\includegraphics[width=\linewidth]{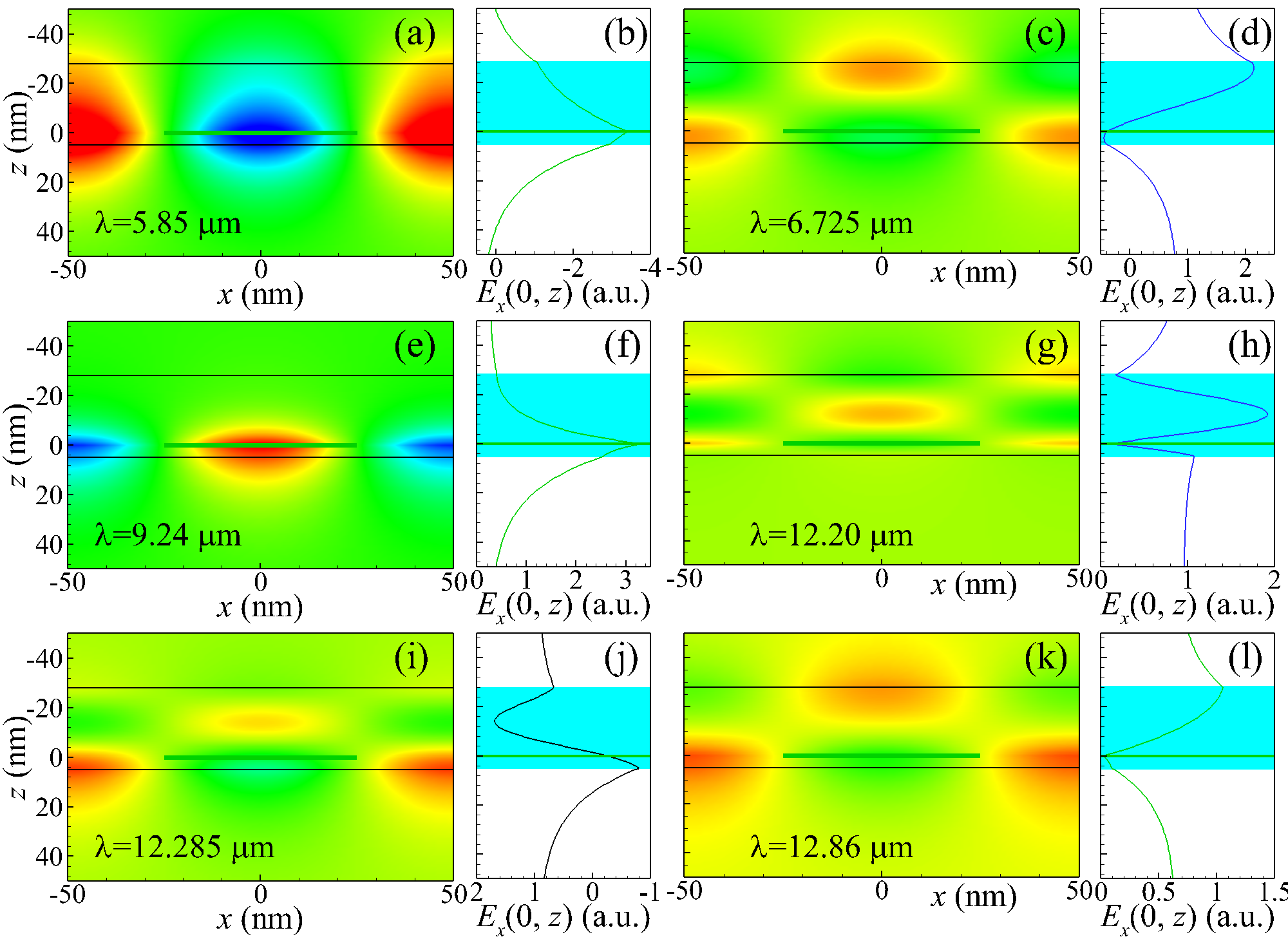}}
		\caption{Spatial distributions of in-plane component of electric field $E_x(x,z)$ [panels (a), (c), (e), (g), (i), (k)] and the same dependencies at the graphene center $E_x(0,z)$ [panels (b), (d), (f), (h), (j), (l)]  at vacuum wavelengths $\lambda=5.85\,\mu$m [panels (a) and (b)], $\lambda=6.725\,\mu$m [panels (c) and (d)], $\lambda=9.24\,\mu$m [panels (e) and (f)], $\lambda=12.2\,\mu$m [panels (g) and (h)], $\lambda=12.285\,\mu$m [panels (i) and (j)], and $\lambda=12.86\,\mu$m [panels (k) and (l)]. Other parameters are the same, as in Fig.\,4 of the main text. In all panels position of graphene strips are depicted by green lines. }
		\label{fig:field-distribution}
	\end{figure}

	Figure \ref{fig:field-distribution} shows spatial distributions of $x$-component of the electric field over one period of the periodic structure (horizontal $x$-axis) at frequencies, corresponding to absorbance maxima in Figure 4 of main text. For the absorbance maximum nearby $\lambda=5.85\,\mu$m electric field distribution [Figure \ref{fig:field-distribution}(a)] is characterized by the minimum in the vicinity of the graphene strip (green horizontal lines) and maximum beyond the graphene (notice that after half-period of temporal oscillations $T=\lambda/(2c)$, where $c$ is the speed of light, these minima and maxima will be inverted). The distribution of electric field along $z$-axis [Figure \ref{fig:field-distribution}(b)], taken at the center of graphene strip $x=0$ demonstrates the maximum of field's absolute value in the vicinity of graphene, thus resembling the eigenmode distribution of Figure 2(e) of main text (plasmonic-like mode). The last fact allows to conclude, that enhanced absorbance happens due to the excitation of plasmonic-like mode.
	
	The field distribution, which corresponds to the absorbance maximum $\lambda=6.725\,\mu$m is characterized by the low field in the vicinity of graphene [see Figure \ref{fig:field-distribution}(c)]. At the same time, electric field distribution along $z$-axis [Figure \ref{fig:field-distribution}(d)] shows one sign variation, which resembles the mode, depicted in Figure 2(f) (waveguide-like mode in upper restrahlen band) of the main text. 
	
	The field distribution, corresponding to absorbance maxima nearby $\lambda=9.24\,\mu$m (between restrahlen bands), shown in Figures \ref{fig:field-distribution}(e) and \ref{fig:field-distribution}(f) is similar to those of absorbance maximum nearby $\lambda=5.85\,\mu$m [Figures \ref{fig:field-distribution}(a) and \ref{fig:field-distribution}(b)], except higher degree of localization of field in the vicinity of graphene for plasmonic-like mode at $\lambda=9.24\,\mu$m.
	
	Electric field distributions, corresponding to absorbance maxima $\lambda=12.2\,\mu$m and $\lambda=12.285\,\mu$m  [depicted in Figures \ref{fig:field-distribution}(g), \ref{fig:field-distribution}(h) and Figures \ref{fig:field-distribution}(i), \ref{fig:field-distribution}(j), respectively] resemble waveguide-like modes in lower restrahlen band, shown in Figures 3(f) and 3(e) of the main text. The situation is more complicated for the field distribution, corresponding to absorbance maximum at $\lambda=12.86\,\mu$m, where field maxima are arranged nearby edges of hBN film [see Figure \ref{fig:field-distribution}(k)]. Except the fact, that absorbance maxima is located beyond the restrahlen bands, the field distribution along $z$-axis, shown in Figure \ref{fig:field-distribution}(l) is not similar to the plasmonic-like mode, shown in Figure 3(i) of the main text.   
	
\end{document}